\documentclass[12pt]{article}

\usepackage{amsmath,amssymb,amsfonts,amsthm}
\usepackage{graphicx}
\usepackage{cite}
\usepackage[all]{xy}

\allowdisplaybreaks[3]

\newcounter{propositiona}
\newcommand{\propositiona}[1]{\refstepcounter{propositiona}
\textbf{Proposition\, \thepropositiona.}\, {\it #1}}
\newcounter{definitiona}
\newcommand{\definitiona}[1]{\refstepcounter{definitiona}
	\textbf{Definition \thedefinitiona.}\, #1}
\newcounter{remarka}
\newcommand{\remarka}[1]{\refstepcounter{remarka}
	\textbf{Remark \theremarka.}\, #1}
\newcounter{examplea}

\newcounter{lemmaa}
\newcommand{\lemmaa}[1]{\refstepcounter{lemmaa}
	\textbf{Lemma \thelemmaa.}\, {\it #1}}
\newcounter{theorema}
\newcommand{\theorema}[1]{\refstepcounter{theorema}
	\textbf{Theorem\, \thetheorema.}\, {\it #1}}
\newcounter{corollarya}
\newcommand{\corollarya}[1]{\refstepcounter{corollarya}
	\textbf{Corollary\, \thecorollarya.}\, {\it #1}}

\begin{document}
	
\large

\thispagestyle{empty}

\begin{center}
{\bf \Large Lagrangian formalism and the intrinsic geometry of PDEs}\\[2.5ex]
{\large \bf Kostya Druzhkov}\\[2.5ex]
Lomonosov Moscow State University, 1 Leninskiye Gory,\\
Main Building, 119991 Moscow, Russia\\[1ex]
\textit{E-mail: konstantin.druzhkov@gmail.com}
\end{center}

\vspace{1.0ex}

\noindent
\textbf{\Large Abstract}\\[1.0ex]
A notion of internal Lagrangian for a system of differential equations is introduced. A spectral sequence related to internal Lagrangians is obtained. A connection between internal Lagrangians and presymplectic structures is investigated. An interpretation of the term $E^{3,\, n-2}_2$ of Vinogradov's $\mathcal{C}$-spectral sequence is given for irreducible gauge theories.\\[1.0ex]

\section{Introduction\\[-5ex]}{\label{Intr}}

\

Stationary-action principles play a significant role in modern theoretical physics. 
Variational principles admit partial description in terms of the intrinsic geometry of the corresponding Euler-Lagrange equations.
Usually, the internal description of stationary-action principles is given in terms of presymplectic structures of differential equations.

\textit{Further, we do not distinguish between variational principles that lead to equations under consideration and variational principles that lead only to differential consequences.} The question of how exactly to distinguish between these two types of variational principles (by means of intrinsic geometry) is non-trivial (see, e.g.,~\cite{GriGri}).

Generally speaking, there are several non-equivalent definitions of presymplectic structures (or presymplectic currents). 
\textit{In what follows, by a presymplectic structure of a differential equation we mean an element of the kernel of the differential $d_1^{\,2,\,n-1}$ from the Vinogradov $\mathcal{C}$-spectral sequence}~\cite{VinKr}.

As it follows from~\cite{Khavk}, a presymplectic structure is related to a variational principle iff it is generated by an exact differential form.
In the general case, the description of variational principles of differential equations in terms of presymplectic structures (as elements of $\ker d^{\,2,\,n-1}_1$) can turn out to be incomplete. Perhaps, a similar situation is possible for conservation laws. More precisely, conservation laws are variational $0$-forms, but in many cases they can be identified with exact variational $1$-forms, i.e., exact elements of the group $E^{1,\, n-1}_1$ of Vinogradov's $\mathcal{C}$-spectral sequence. Variational $1$-forms themselves are completely characterized by cosymmetries~\cite{VinKr} (or characteristics). This approach works when the corresponding differential $d_1^{\,0,\,n-1}$ is monomorphic. Otherwise, the description of conservation laws in terms of variational $1$-forms is incomplete and there exist nontrivial conservation laws with trivial cosymmetries (however, the author does not know examples of such conservation laws). Speaking of variational principles, the role of presymplectic structures is analogous to the role of variational $1$-forms for conservation laws.

In the paper, we introduce internal Lagrangians as elements of cohomology of some complex on a differential equation (namely, the complex~\eqref{mainc}). This is an attempt to answer the following question. \textit{Where exactly does a differential equation contain information about its variational nature?} Thus the notion of internal Lagrangian differs from the notion of intrinsic Lagrangian introduced in~\cite{Grigoriev}. 
The connection between internal Lagrangians and presymplectic structures is based on a spectral sequence analogous to Vinogradov's $\mathcal{C}$-spectral sequence. The results obtained allow us to get a condition under which internal Lagrangians can be identified with presymplectic structures.

The paper is organized as follows.
In section~\ref{Def} we introduce basic notations and recall some facts from the geometry of differential equations. 
In section~\ref{IntL} we describe the notion of internal Lagrangian. 
Section~\ref{St} is devoted to the relation between internal Lagrangians and stationary-action principles. There we show that each internal Lagrangian of a system of differential equations is generated by some action such that all solutions of the system of equations under consideration are its stationary points. In section~\ref{Sp} we introduce a spectral sequence that allows us to obtain a relation between internal Lagrangians and presymplectic structures. We also show there that if an equation admits a
compatibility complex of length three, the incompleteness of intrinsic description of variational principles by presymplectic structures is determined by the term $E^{3,\, n-2}_2$ of the Vinogradov $\mathcal{C}$-spectral sequence. These results are applicable to irreducible gauge theories.
In section~\ref{Dis} we discuss the above-mentioned results and apply them to Abelian $p$-form theories.

\section{Basic notations and definitions\\[-5 ex]}\label{Def} 
\

Let us recall some facts from the geometry of differential equations.\\[-5ex]

\subsection{Jets (see, e.g.,~\cite{VinKr})\\[-6 ex]}

\

Let $\pi\colon E\to M$ be a locally trivial smooth vector
bundle over a smooth manifold $M$, $\mathrm{dim}\, M = n$, 
$\mathrm{dim}\, E = n + m$. The bundle $\pi$ allows us to introduce the corresponding jet bundles
\begin{align*}
\xymatrix{
\ldots \ar[r] & J^3(\pi) \ar[r]^-{\pi_{3, 2}} & J^2(\pi) \ar[r]^-{\pi_{2, 1}} & J^1(\pi) \ar[r]^-{\pi_{1, 0}} & J^0(\pi) = E \ar[r]^-\pi & M
}
\end{align*}
and the inverse limit $J^{\infty}(\pi)$ of this chain. Denote by $\mathcal{F}(\pi)$ the algebra of smooth functions on $J^{\infty}(\pi)$.

Suppose $U\subset M$ is a coordinate neighborhood such that the bundle $\pi$
becomes trivial over $U$. Choose local coordinates $x^1$, \ldots, $x^n$ in $U$ and $u^1$, \ldots, $u^m$  
along the fibers of $\pi$ over $U$. It is convenient to introduce
multi-index $\alpha = (\alpha_1,\ldots, \alpha_n)$, where all $\alpha_i$ are non-negative integers.
We denote the corresponding adapted local coordinates on $J^{\infty}(\pi)$ by $u^i_{\alpha}$.

The main structure on $J^{\infty}(\pi)$ is a Cartan distribution $\mathcal{C}$. 
The Cartan distribution is spanned by the total derivatives (using the summation convention)
$$
D_{x^i} = \partial_{x^i} + u^k_{\alpha + 1_i}\partial_{u^k_{\alpha}}\,,\qquad i = 1,\ldots, n.
$$
Here $\alpha + 1_i = (\alpha_1,\ldots,\alpha_i + 1, \ldots, \alpha_n)$.

The Cartan distribution determines an ideal $\mathcal{C}\Lambda^*(\pi)\subset \Lambda^*(\pi)$ 
of the algebra of differential forms on $J^{\infty}(\pi)$.
The ideal $\mathcal{C}\Lambda^*(\pi)$ consists of Cartan forms, i.e., differential forms vanishing on the Cartan distribution.
A Cartan $1$-form $\omega\in\mathcal{C}\Lambda^1(\pi)$ can be written as a finite sum
$$
\omega = \omega_i^{\alpha}\theta^i_{\alpha}\,,\qquad \theta^i_{\alpha} = du^i_{\alpha} - u^i_{\alpha + 1_k}dx^k
$$
in adapted local coordinates. Here the coefficients $\omega_i^{\alpha}$ are smooth functions defined on the coordinate domain.

Powers of the ideal $\mathcal{C}\Lambda^*(\pi)$ are stable with respect to the de Rham differential, i.e., $d(\mathcal{C}^p\Lambda^*(\pi))\subset \mathcal{C}^p\Lambda^*(\pi)$, where
$$
\mathcal{C}^p\Lambda^*(\pi) = \underbrace{\mathcal{C}\Lambda^1(\pi)\wedge\ldots\wedge \mathcal{C}\Lambda^1(\pi)}_{p}\wedge\, \Lambda^*(\pi)\,.
$$
Thus the de Rham complex admits the filtration
$$
\Lambda^*(\pi)\supset \mathcal{C}\Lambda^*(\pi)\supset \mathcal{C}^2\Lambda^*(\pi)\supset \ldots
$$
The corresponding spectral sequence $(E^{p,\, q}_r(\pi), d_r)$ is the Vinogradov $\mathcal{C}$-spectral sequence. Here $E^{p,\, q}_1(\pi) = 0$ for $p > 0$, $q \leqslant n-1$.

Let $\varkappa(\pi) = \Gamma(\pi^*_{\infty}(\pi))$ be the $\mathcal{F}(\pi)$-module
of sections of the corresponding pullback. Here $\pi_{\infty}\colon J^{\infty}(\pi)\to M$ is a natural projection.
If $\varphi\in \varkappa(\pi)$, there is the corresponding evolutionary vector field on $J^{\infty}(\pi)$
$$
E_{\varphi} = D_{\alpha}(\varphi^i)\partial_{u^i_{\alpha}}\,,
$$
where $\varphi^1$, \ldots, $\varphi^m$ are components of $\varphi$ in adapted local coordinates, $D_{\alpha}$ is the composition $D_{\alpha} = D_{x^1}^{\ \alpha_1}\circ\ldots\circ D_{x^n}^{\ \alpha_n}$. Evolutionary vector fields are infinitesimal symmetries of $J^{\infty}(\pi)$, i.e., $\mathcal{L}_{E_{\varphi}}(\mathcal{C}\Lambda^*(\pi))\subset \mathcal{C}\Lambda^*(\pi)$. Here $\mathcal{L}_{E_{\varphi}}$ is the corresponding Lie derivative.

Cartan forms allow us to consider horizontal $k$-forms on $J^{\infty}(\pi)$:
$$
\Lambda^k_h(\pi) = \Lambda^k(\pi)/\mathcal{C}\Lambda^k(\pi)\,.
$$
A differential $k$-form $\omega\in\Lambda^k(\pi)$ generates the horizontal $k$-form $[\omega]_h = \omega + \mathcal{C}\Lambda^k(\pi)$.
The infinite jet bundle $\pi_{\infty}\colon J^{\infty}(\pi) \to M$ admits the decomposition 
$$
\Lambda^1(\pi) = \mathcal{C}\Lambda^1(\pi) \oplus \mathcal{F}(\pi)\!\cdot\!\pi^*_{\infty}(\Lambda^1(M))\,.
$$
Thus the module of horizontal $k$-forms $\Lambda^k_h(\pi)$ can be identified with $\mathcal{F}(\pi)\cdot \pi^*_{\infty}(\Lambda^k(M))$.
The de Rham differential $d$ induces a horizontal differential
\begin{equation*}
d_h\colon \Lambda^{*}_h(\pi)\to \Lambda^{*}_h(\pi)\,,\qquad d_h[\omega]_h = [d\omega]_h\,.
\end{equation*}

Let $\xi\colon N\to M$ be a locally trivial smooth vector
bundle. Suppose $P(\pi)$ is a module of sections of the pullback $\pi^*_{\infty}(\xi)$:
$$
P(\pi) = \Gamma(\pi^*_{\infty}(\xi))\,.
$$
Then we can consider the adjoint module
\begin{equation*}
\widehat{P}(\pi) = \mathrm{Hom}_{\mathcal{F}(\pi)}(P(\pi), \Lambda^n_h(\pi))\,.
\end{equation*}

By a $\mathcal{C}$-differential operator we will mean an operator in total derivatives. Suppose $Q_1$, $Q_2$ are modules of sections of some vector bundles over $J^{\infty}(\pi)$. Denote by $\mathcal{C}(Q_1, Q_2)$ 
a module of $\mathcal{C}$-differential operators
from $Q_1$ to $Q_2$. If $\Delta\in \mathcal{C}(\varkappa(\pi), \mathcal{F}(\pi))$, $\varphi\in \varkappa(\pi)$, 
we obtain the relation
$$
\Delta(\varphi) = \Delta^{\alpha}_i D_{\alpha}(\varphi^i) = i_{E_{\varphi}}\,\Delta^{\alpha}_i\, \theta^i_{\alpha}\,.
$$
In fact, we can identify the $\mathcal{F}(\pi)$-modules $\mathcal{C}(\varkappa(\pi), \mathcal{F}(\pi))$ and $\mathcal{C}\Lambda^1(\pi)$ because of this relation.

Consider the complex
$$
\xymatrix
{
0\ar[r] &\mathcal{C}(\varkappa(\pi), \mathcal{F}(\pi))\ar[r]^-{d_h\circ} &\ldots \ar[r]^-{d_h\circ} 
&\mathcal{C}(\varkappa(\pi), \Lambda^{n-1}_h(\pi))\ar[r]^-{d_h\circ} 
&\mathcal{C}(\varkappa(\pi), \Lambda^n_h(\pi))\ar[r] &0\,.
}
$$
It is important for us that $n$-th cohomology group of the complex is $\widehat{\varkappa}(\pi)$ (other cohomology groups are trivial) 
and each operator $\Delta\in \mathcal{C}(\varkappa(\pi), \Lambda^n_h(\pi))$ 
can be decomposed to a sum
\begin{align}
\Delta = d_h\circ\Delta_1 + \mu\,,
\label{deco}
\end{align}
where $\mu\in\widehat{\varkappa}(\pi)$ is uniquely determined.
Similarly, a $\mathcal{C}$-differential operator $\nabla\colon P(\pi)\to \mathcal{C}(\varkappa(\pi), \Lambda^n_h(\pi))$ admits a decomposition of the form
\begin{align}
\nabla(G) = d_h\circ\nabla_1(G) + A(G)\qquad \forall\, G\in P(\pi)\,,
\label{decom}
\end{align}
where $A\in \mathcal{C}(P(\pi), \widehat{\varkappa}(\pi))$ is also uniquely determined.

By a $\mathcal{C}$-differential operator $B\colon P(\pi)\times\varkappa(\pi)\to \Lambda^k_h(\pi)$ we will mean a $\mathcal{C}$-differential operator 
$B'\colon P(\pi)\to\mathcal{C}(\varkappa(\pi), \Lambda^k_h(\pi))$ such that
$$
B(G, \varphi) = B'(G)(\varphi)
$$
for all $G\in P(\pi)$, $\varphi\in\varkappa(\pi)$.

Denote by $\mathrm{E}$ the Euler operator (variational derivative), $\mathrm{E}\colon \Lambda^n_h(\pi)\to \widehat{\varkappa}(\pi)$. Suppose $L$ is a differential $n$-form, $L\in\Lambda^n(\pi)$. If $L = L_{1...n}\, dx^1\wedge\ldots\wedge dx^n$, we get
\begin{align*}
&\langle \mathrm{E}[L]_h, \varphi\rangle = (-1)^{|\alpha|}D_{\alpha}\Big(\dfrac{\partial L_{1...n}}{\partial u^i_{\alpha}}\Big)\, \varphi^i dx^1\wedge\ldots\wedge dx^n\,,\\
&\mathrm{E}[L]_h = (-1)^{|\alpha|}D_{\alpha}\Big(\dfrac{\partial L_{1...n}}{\partial u^i_{\alpha}}\Big)\, \theta^i_0\wedge dx^1\wedge\ldots\wedge dx^n\,.
\end{align*}
Here $\langle \cdot , \cdot \rangle$ is the natural pairing between the module and its adjoint; $|\alpha| = \sum_{i=1}^{n}\alpha_i$.

The Noether formula links the Lie derivative $\mathcal{L}_{E_{\varphi}}[L]_h$ of a horizontal $n$-form $[L]_h$ and the variational derivative $\mathrm{E}[L]_h$. Namely,
there exists a differential form $\omega_L\in \mathcal{C}\Lambda^{n}(\pi)$ such that
\begin{align}
\mathcal{L}_{E_{\varphi}}[L]_h = \langle \mathrm{E}[L]_h, \varphi \rangle + d_h [i_{E_{\varphi}}\omega_L]_h\,.
\label{Noeth}
\end{align}

\subsection{Differential equations\\[-5 ex]}

\

Let $F$ be a (smooth) section of the bundle $\pi^*_{k}(\xi)$ such that $\{F = 0\}\subset J^{k}(\pi)$ is an embedded submanifold. We also assume that for each $p\in \{F = 0\}$ the differentials of coordinate functions $dF^i|_{p}$ are linearly independent. By infinite prolongation of the differential equation $F = 0$ we will mean the subset $\mathcal{E}\subset J^{\infty}(\pi)$ defined by the following infinite system of equations
\begin{align*}
\quad D_{\alpha}(F^i) = 0\,,\qquad |\alpha| \geqslant 0\,.
\end{align*}\\[-2 ex]
\remarka{We do not require that the number of equations of the form $F^i = 0$ coincide with the number of unknowns.}\\[-1 ex]

Henceforth we always assume that $\pi_{\infty}(\mathcal{E}) = M$. By $\mathcal{F}(\mathcal{E})$ we denote the algebra of smooth functions on $\mathcal{E}$: 
$$
\mathcal{F}(\mathcal{E}) = \mathcal{F}(\pi)|_{\mathcal{E}} = \mathcal{F}(\pi)/I\,,
$$
where $I$ is the ideal of the system $\mathcal{E}\subset J^{\infty}(\pi)$. The algebra of smooth functions $\mathcal{F}(\mathcal{E})$ produces the algebra of differential forms $\Lambda^*(\mathcal{E})$. Here $\Lambda^*(\mathcal{E}) = \Lambda^*(\pi)|_{\mathcal{E}}$. The Cartan distribution on $J^{\infty}(\pi)$ can be restricted to $\mathcal{E}$. Thus we obtain an ideal $\mathcal{C}\Lambda^*(\mathcal{E})\subset \Lambda^*(\mathcal{E})$. Here $\mathcal{C}\Lambda^*(\mathcal{E})$ consists of differential forms vanishing on the Cartan distribution (on $\mathcal{E}$).\\[1ex]
\definitiona{An infinitely prolonged system of equations $\mathcal{E}$ is \textit{regular} if\\
1) a function $f\in \mathcal{F}(\pi)$ vanishes on $\mathcal{E}$ (i.e., $f\in I$) iff there exists a $\mathcal{C}$-differential operator $\Delta\colon P(\pi)\to \mathcal{F}(\pi)$} such that $f = \Delta(F)$;\\
2) $\mathcal{C}\Lambda^*(\mathcal{E}) = \mathcal{C}\Lambda^*(\pi)|_{\mathcal{E}} = \mathcal{C}\Lambda^*(\pi)/\big(I\cdot \mathcal{C}\Lambda^*(\pi) + dI\wedge \mathcal{C}\Lambda^*(\pi)\big)$.\\[1 ex]
\textit{In what follows, we 
assume that de Rham cohomology groups $H^i_{dR}(\mathcal{E})$ are trivial for $i > 0$.} These conditions are not restrictive.
The case of nontrivial de Rham cohomology does not lead to essential complications. In this case, one can factorize
all the cohomology considered below by their subgroups generated by the de Rham cocycles.

\section{\label{IntL} Internal Lagrangians}

To introduce notations, we begin with a simple\\[1 ex]
\lemmaa{\label{l1} Suppose $L\in\Lambda^n(\pi)$ is a differential $n$-form (on jets). 
Then there exists a Cartan $n$-form $\omega_L\in \mathcal{C}\Lambda^{n}(\pi)$ such that the relation
\begin{align}
\langle \mathrm{E}[L]_h, \varphi\rangle = [i_{E_{\varphi}}d(L + \omega_L)]_h
\label{map}
\end{align}
holds for all $\varphi\in\varkappa(\pi)$.
}\\[1 ex]
\textbf{Proof.}
According to Noether's formula, there exists a differential form $\widetilde{\omega}_L\in \mathcal{C}\Lambda^{n}(\pi)$ such that
$$
\mathcal{L}_{E_{\varphi}}[L]_h = \langle \mathrm{E}[L]_h, \varphi \rangle + d_h [i_{E_{\varphi}}\widetilde{\omega}_L]_h\,.
$$
The horizontal $n$-form $[L]_h$ is generated by a unique $n$-form $\widetilde{L}\in\mathcal{F}(\pi)\cdot\pi_{\infty}^*(\Lambda^n(M))$. Then $L - \widetilde{L} \in\mathcal{C}\Lambda^n(\pi)$ and 
$$
\mathcal{L}_{E_{\varphi}}[L]_h = [\mathcal{L}_{E_{\varphi}}\widetilde{L} + \mathcal{L}_{E_{\varphi}}(L - \widetilde{L})]_h = [\mathcal{L}_{E_{\varphi}}\widetilde{L}]_h\,.
$$
Since $i_{E_{\varphi}}\widetilde{L} = 0$ and 
$d_h [i_{E_{\varphi}}\widetilde{\omega}_L]_h = [\mathcal{L}_{E_{\varphi}}\widetilde{\omega}_L - i_{E_{\varphi}}d\widetilde{\omega}_L]_h = -[i_{E_{\varphi}}d\widetilde{\omega}_L]_h$,
we obtain the relation
$$
\langle \mathrm{E}[L]_h, \varphi\rangle = [i_{E_{\varphi}}d(\widetilde{L} + \widetilde{\omega}_L)]_h\,.
$$
Thus one can put $\omega_L = \widetilde{L} - L + \widetilde{\omega}_L \in \mathcal{C}\Lambda^n(\pi)$.\\[1 ex]
\remarka{The idea of relating a Lagrangian $[L]_h$ to forms like $L + \omega_L$ from Lemma~\ref{l1} is by no means new (see, e.g.,~\cite{Zuc}, \cite{KruSan}). Apparently, a systematic study of the cohomology classes associated with them has not been carried out before.}\\[-1 ex]

Now let us consider a system of differential equations $\{F = 0\}\subset J^{k}(\pi)$ and its infinite prolongation 
$\mathcal{E}\subset J^{\infty}(\pi)$.\\[1 ex]
\lemmaa{\label{lemsep}The formula~\eqref{map} determines a homomorphism of vector spaces 
$$
\{[L]_h\in\Lambda^n_h(\pi)\colon\, \mathrm{E}[L]_h|_{\mathcal{E}} = 0\}\to 
\dfrac{\{\omega\in\Lambda^n(\mathcal{E})\colon d\omega\in \mathcal{C}^2\Lambda^{n+1}(\mathcal{E})\}}
{\mathcal{C}^2\Lambda^n(\mathcal{E}) + d(\mathcal{C}\Lambda^{n-1}(\mathcal{E}))}\,.
$$}\\
\textbf{Proof.} Assume that the variational derivative $\mathrm{E}[L]_h$ vanishes on the system~$\mathcal{E}$. Then the formula~\eqref{map} allows us to conclude that 
$$
d(L + \omega_L)|_{\mathcal{E}} \in \mathcal{C}^2\Lambda^{n+1}(\mathcal{E})\,.
$$
If $L'\in\Lambda^n(\pi)$ and $\omega_L'\in\mathcal{C}\Lambda^n(\pi)$ are differential forms such that $[L]_h = [L']_h$ and
$$
\langle \mathrm{E}[L]_h, \varphi\rangle = [i_{E_{\varphi}}d(L' + \omega_L')]_h\,,
$$
then
$$
[i_{E_{\varphi}}d(L + \omega_L - L' - \omega_L')]_h = 0\,.
$$
Thus $d(L - L' + \omega_L - \omega_L')\in\mathcal{C}^2\Lambda^{n+1}(\pi)$. However, the group $E^{1,\, n-1}_1(\pi)$ of the Vinogradov $\mathcal{C}$-spectral sequence is trivial. Hence, 
$$
L - L' + \omega_L - \omega_L'\in \mathcal{C}^2\Lambda^n(\pi) + d(\mathcal{C}\Lambda^{n-1}(\pi))
$$
and the differential form $(L' + \omega_L')|_{\mathcal{E}}$ produces the same element of the desired quotient space.\\[1 ex]
\remarka{\label{rem2} In some situations it may be important to consider Lagrangians on jets precisely as separate horizontal $n$-forms (instead of considering up to additive exact horizontal forms). In this case, the internal description of such Lagrangians is given by the quotient space from Lemma~\ref{lemsep}.\\[-1 ex]}

In a similar way we get the following\\[1 ex]
\propositiona{Let $[L]_h$ be a horizontal $n$-form such that $\mathrm{E}[L]_h$ vanishes on an infinitely prolonged system of differential equations~$\mathcal{E}$. Suppose $[\eta]_h$ is a horizontal $n$-form such that $[L]_h - [\eta]_h \in \mathrm{im}\, d_h$. Then $[L]_h$ and $[\eta]_h$ produce the same elements of
the $(n-1)^{\text{th}}$ cohomology of the complex
\begin{align}
\ldots\! \overset{d_{n-3}}{\to}\Lambda^{n-1}(\mathcal{E})/\mathcal{C}^2\Lambda^{n-1}(\mathcal{E})\overset{d_{n-2}}{\to} \Lambda^{n}(\mathcal{E})/\mathcal{C}^2\Lambda^{n}(\mathcal{E})\overset{d_{n-1}}{\to} \Lambda^{n+1}(\mathcal{E})/\mathcal{C}^2\Lambda^{n+1}(\mathcal{E})\to 0.
\label{mainc}
\end{align}}
Here the differentials $d_i$ are induced by the de Rham differential $d$.\\[-1 ex]

\textit{In the sequel, we will refer to the elements of the $(n-1)^{\text{th}}$ cohomology of the complex~\eqref{mainc} as internal Lagrangians.} 
Such a description of internal Lagrangians of an infinitely prolonged system $\mathcal{E}$ has two important advantages. 
Firstly, each variational principle uniquely determines an internal Lagrangian.
\textit{The approach that defines a presymplectic structure as a single differential form does not have this advantage.} Secondly, such a description allows one to relate a group of internal Lagrangians to Vinogradov's $\mathcal{C}$-spectral sequence (see section~\ref{Sp}).\\[1ex]
\remarka{In the language of the variational bicomplex~\cite{And}, the analogue of the quotient space from Lemma~\ref{lemsep} is
$$
\dfrac{\{\omega\in \Omega^{n, 0}\oplus\Omega^{n-1, 1}\colon\ (d_V\circ \mathrm{pr}_{n,0} + d_H\circ \mathrm{pr}_{n-1,1})(\omega) = 0\}}{d_H\Omega^{n-2,1}}\,.
$$
Here elements of $\Omega^{i, j}$ are differentail forms (on $\mathcal{E}$) of horizontal degree $i$ and vertical degree $j$;\, $\mathrm{pr}_{n,0}\colon \Omega^{n, 0}\oplus\Omega^{n-1, 1}\to \Omega^{n, 0}$ and $\mathrm{pr}_{n-1,1}\colon \Omega^{n, 0}\oplus\Omega^{n-1, 1}\to \Omega^{n-1, 1}$ are natural projections. The analogue of the $(n-1)^{\text{th}}$ cohomology of the complex~\eqref{mainc} is
$$
\dfrac{\{\omega\in \Omega^{n, 0}\oplus\Omega^{n-1, 1}\colon\ (d_V\circ \mathrm{pr}_{n,0} + d_H\circ \mathrm{pr}_{n-1,1})(\omega) = 0\}}
{(d_H + d_V)\Omega^{n-1,0} + d_H\Omega^{n-2,1}}\,.
$$
}

\section{\label{St} Stationary-action principles from internal Lagrangians}

Let us consider a system of differential equations $\{F = 0\}\subset J^{k}(\pi)$ and its infinite prolongation $\mathcal{E}\subset J^{\infty}(\pi)$. We assume that the system of equations $\mathcal{E}$ is regular.\\[1.5 ex]
\lemmaa{\label{L1}Suppose $l\in\Lambda^n(\mathcal{E})$ is a differential $n$-form  such that $dl \in \mathcal{C}^2\Lambda^{n+1}(\mathcal{E})$. Then there is a differential $n$-form $L\in\Lambda^n(\pi)$ such that
$$
L|_{\mathcal{E}} = l\,,\qquad dL \in \mathcal{C}^2\Lambda^{n+1}(\pi) + I\cdot \Lambda^{n+1}(\pi)\,.
$$
}\\[-6 ex]

\noindent
\textbf{Proof.} Since the system~$\mathcal{E}$ is regular, there exist differential forms $a\in \Lambda^{n}(\pi)$, $b\in \mathcal{C}^2\Lambda^{n+1}(\pi)$ such that
$$
a|_{\mathcal{E}} = l\,,\qquad b|_{\mathcal{E}} = dl\,.
$$
Hence, $da - b\in I\cdot \Lambda^{n+1}(\pi) + dI\wedge \Lambda^{n}(\pi)$ and there are functions $f_i\in I$ and differential forms $\omega^i\in \Lambda^{n}(\pi)$ such that
$$
da - b - df_i\wedge \omega^i\in I\cdot \Lambda^{n+1}(\pi)\,.
$$
Thus $d(a - f_i\,\omega^i) - b \in I\cdot \Lambda^{n+1}(\pi)$ and one can put $L = a - f_i\,\omega^i$.\\[1ex]
\theorema{\label{theor}
Let $l\in\Lambda^n(\mathcal{E})$ be a differential $n$-form such that $dl\in \mathcal{C}^2\Lambda^{n+1}(\mathcal{E})$.
Suppose $L$ is an extension of $l$ from Lemma~\ref{L1}. Then\\
1) the variational derivative $\mathrm{E}[L]_h$ vanishes on $\mathcal{E}$;\\
2) $l$ and $[L]_h$ produce the same elements of the quotient space from Lemma~\ref{lemsep}.}\\[1ex]
\textbf{Proof.}\\ 
1) Since $dL \in \mathcal{C}^2\Lambda^{n+1}(\pi) + I\cdot \Lambda^{n+1}(\pi)$, there exists a $\mathcal{C}$-differential operator 
$\nabla\colon P(\pi)\times \varkappa(\pi)\to \Lambda_h^{n}(\pi)$ such that
\begin{align}
\nabla(F, \varphi) = [i_{E_{\varphi}}dL]_h\,.
\label{nab}
\end{align}
Therefore, we obtain the relation
\begin{align*}
\mathcal{L}_{E_{\varphi}}[L]_h - d_h[i_{E_{\varphi}}L]_h = [\mathcal{L}_{E_{\varphi}}L - di_{E_{\varphi}}L]_h = \nabla(F, \varphi)\,.
\end{align*}
The operator $\nabla$ can be decomposed to a sum of the form~\eqref{decom}
\begin{align}
\nabla(G, \cdot) = d_h\circ\nabla_1(G, \cdot) + \langle A(G), \cdot\rangle\qquad \forall\, G\in P(\pi)\,.
\label{decom1}
\end{align}
Here the operator $A\colon P(\pi) \to \widehat{\varkappa}(\pi)$ is uniquely determined. Hence, 
\begin{align*}
\mathcal{L}_{E_{\varphi}}[L]_h = \langle A(F), \varphi\rangle + d_h\big([i_{E_{\varphi}}L]_h + \nabla_1(F, \varphi)\big)
\end{align*}
and we obtain the decomposition of the form~\eqref{deco} for the operator $\Delta\colon \varphi\mapsto \mathcal{L}_{E_{\varphi}}[L]_h$.
However, the Noether formula~\eqref{Noeth} gives us a decomposition of a similar form.
Thus
\begin{align}
\mathrm{E}[L]_h = A(F)
\label{iden}
\end{align}
because of uniqueness of $\mu$ in~\eqref{deco}.\\
2) The operator $\nabla_1(F, \cdot)$ can be represented by a form $\omega_L\in I\cdot\mathcal{C}\Lambda^n(\pi)$:
$$
\nabla_1(F, \varphi) = [i_{E_{\varphi}}\omega_L]_h\,.
$$
Therefore, the relations~\eqref{nab}, \eqref{decom1} and \eqref{iden} imply
\begin{align*}
d_h[i_{E_{\varphi}}\omega_L]_h + \langle \mathrm{E}[L]_h, \varphi\rangle = [i_{E_{\varphi}}dL]_h\,.
\end{align*}
Besides, we have $d_h[i_{E_{\varphi}}\omega_L]_h = [\mathcal{L}_{E_{\varphi}}\omega_L - i_{E_{\varphi}}d\omega_L]_h = -[i_{E_{\varphi}}d\omega_L]_h$.
Thus
$$
\langle \mathrm{E}[L]_h, \varphi\rangle = [i_{E_{\varphi}}d(L + \omega_L)]_h
$$
and the corresponding element of the quotient space from Lemma~\ref{lemsep} is generated by the form $(L + \omega_L)|_{\mathcal{E}} = l$.\\[-1ex]

Let us note that a theorem close in meaning to Theorem~\ref{theor} is given in~\cite{Khavk} (Theorem 3). Theorem~\ref{theor} is consistent with the description of Lagrangians (on jets) as separate horizontal $n$-forms (as in Remark~\ref{rem2}).

\section{\label{Sp} Spectral sequence for Lagrangian formalism}

Let us consider the de Rham complex for an infinitely prolonged system of differential equations $\mathcal{E}$
\begin{align*}
\xymatrix
{
0\ar[r] &\mathcal{F}(\mathcal{E})\ar[r]^d &\Lambda^{1}(\mathcal{E})\ar[r]^d &\Lambda^{2}(\mathcal{E})\ar[r]^d &\Lambda^{3}(\mathcal{E})\ar[r]^d &\ldots
}
\end{align*}
Since all ideals of the form $\mathcal{C}^{p}\Lambda^*(\mathcal{E})$ are stable with respect to the de Rham differential, we can consider the
filtration 
\begin{align}
\Lambda^*(\mathcal{E})\supset \mathcal{C}^2\Lambda^{*}(\mathcal{E})\supset \mathcal{C}^3\Lambda^{*}(\mathcal{E})\supset \mathcal{C}^4\Lambda^{*}(\mathcal{E})\supset \ldots
\label{filtr}
\end{align}
This filtration is finite in each degree because of the identities $\mathcal{C}^{k+1}\Lambda^k(\mathcal{E}) = 0$.\\[1.5 ex]
\remarka{The Vinogradov $\mathcal{C}$-spectral sequence $(E^{p, q}_r(\mathcal{E}), d^{\,p,q}_r)$ is produced by the filtration
\begin{align*}
\Lambda^*(\mathcal{E})\supset \mathcal{C}\Lambda^{*}(\mathcal{E})\supset \mathcal{C}^2\Lambda^{*}(\mathcal{E})\supset \mathcal{C}^3\Lambda^{*}(\mathcal{E})\supset \ldots
\end{align*}
}

The filtration~\eqref{filtr} determines the corresponding spectral sequence $(\widetilde{E}^{\,p, q}_r(\mathcal{E}), \tilde{d}^{\,p,q}_r)$, which converges to the de Rham cohomology and is related to Vinogradov's $\mathcal{C}$-spectral sequence $(E^{p, q}_r(\mathcal{E}), d^{\,p,q}_r)$. Namely, the relation between these spectral sequences is based on the identities (for $p \geqslant 1$)
$$
\widetilde{E}^{p,\, q}_0(\mathcal{E}) = E^{p+1,\, q}_0(\mathcal{E})\,,\qquad \tilde{d}_0^{\,p ,\, q} = d_0^{\,p+1 ,\, q}\,.
$$
Besides, the terms $\widetilde{E}^{\,0,\, q}_0(\mathcal{E})$ form the complex~\eqref{mainc}:
$$
\widetilde{E}^{\,0,\, q}_0 = \Lambda^{q+1}(\mathcal{E})/\mathcal{C}^2\Lambda^{q+1}(\mathcal{E})
$$
Thus, we have the differential 
\begin{align*}
\tilde{d}_1^{\,0,\, n-1}\colon \widetilde{E}^{\,0,\, n-1}_1(\mathcal{E})\to E^{2,\, n-1}_1(\mathcal{E})\,,
\end{align*}
which actually sends internal Lagrangians to presymplectic structures.\\[1ex]
\definitiona{
A \textit{hidden Lagrangian} of an infinitely prolonged system of differential equations is an element of the group $\ker \tilde{d}_1^{\,0,\, n-1}$.\\[1ex]}
Hidden Lagrangians are invisible to presymplectic structures. Thus a presymplectic structure determines internal Lagrangian up to a hidden one.\\[1ex]
\remarka{
There is the analogy between internal Lagrangians and conservation laws. Namely, $\widetilde{E}^{\,0,\, n-1}_1(\mathcal{E})$ is the group of internal Lagrangians of a system $\mathcal{E}$, while $E^{\,0,\, n-1}_1(\mathcal{E})$ is the group of its conservation laws.\\[-1.5 ex]}

The first page of the spectral sequence $(\widetilde{E}^{\,p,\, q}_r(\mathcal{E}), \tilde{d}^{\,p,q}_r)$ reads
\begin{align*}
\xymatrixcolsep{3pc}
\xymatrix @R = 0.4 pc
{
\ \widetilde{E}^{\,0,\, n}_1(\mathcal{E})\ \ar[r]^{\tilde{d}_1^{\,0, n}} &\ E^{2,\, n}_1(\mathcal{E})\ \ar[r]^{d_1^{\,2, n}} &\ E^{3,\, n}_1(\mathcal{E})\ \ar[r]^{d_1^{\,3, n}} &\ E^{4,\, n}_1(\mathcal{E})\ar[r]\ &\ \ldots\\
\widetilde{E}^{\,0,\, n-1}_1(\mathcal{E})\ar[r]^-{\tilde{d}_1^{\,0, n-1}} &E^{2,\, n-1}_1(\mathcal{E})\ar[r]^{d_1^{\,2, n-1}} &E^{3,\, n-1}_1(\mathcal{E})\ar[r]^{d_1^{\,3, n-1}} &E^{4,\, n-1}_1(\mathcal{E})\ar[r] &\ \ldots\\
\widetilde{E}^{\,0,\, n-2}_1(\mathcal{E})\ar[r]^-{\tilde{d}_1^{\,0, n-2}} &E^{2,\, n-2}_1(\mathcal{E})\ar[r]^{d_1^{\,2, n-2}} &E^{3,\, n-2}_1(\mathcal{E})\ar[r]^{d_1^{\,3, n-2}} &E^{4,\, n-2}_1(\mathcal{E})\ar[r] &\ \ldots\\
\widetilde{E}^{\,0,\, n-3}_1(\mathcal{E})\ar[r]^-{\tilde{d}_1^{\,0, n-3}} &E^{2,\, n-3}_1(\mathcal{E})\ar[r]^{d_1^{\,2, n-3}} &E^{3,\, n-3}_1(\mathcal{E})\ar[r]^{d_1^{\,3, n-3}} &E^{4,\, n-3}_1(\mathcal{E})\ar[r] &\ \ldots\\
&& \ldots\\
\ \widetilde{E}^{\,0,\, 0}_1(\mathcal{E})\ \ar[r]^-{\tilde{d}_1^{\,0, 0}} &\ E^{2,\, 0}_1(\mathcal{E})\ \ar[r]^{d_1^{\,2, 0}} &\ E^{3,\, 0}_1(\mathcal{E})\ \ar[r]^{d_1^{\,3, 0}} &\ E^{4,\, 0}_1(\mathcal{E})\ar[r]\ &\ \ldots
}
\end{align*}
Here for $i\geqslant 0$ we have $\widetilde{E}^{1+i,\, *}_1(\mathcal{E}) = E^{2+i,\, *}_1(\mathcal{E})$ and hence
$$
\widetilde{E}^{2+i,\, *}_2(\mathcal{E}) = E^{3+i,\, *}_2(\mathcal{E}),\quad \widetilde{E}^{4+i,\, *}_3(\mathcal{E}) = E^{5+i,\, *}_3(\mathcal{E}),
\quad \widetilde{E}^{7+i,\, *}_4(\mathcal{E}) = E^{8+i,\, *}_4(\mathcal{E}), \quad \ldots
$$
Thus if $a_r$ is a sequence such that $a_1 = 2$, $a_{r+1} = a_r + r$, we obtain 
$$
\widetilde{E}^{a_r - 1 + i,\, *}_r(\mathcal{E}) = E^{a_r + i,\, *}_r(\mathcal{E})\qquad \text{for}\quad r\geqslant 1,\ i\geqslant 0.
$$

If the groups $\widetilde{E}^{r,\,n-r}_r(\mathcal{E})$ are trivial for $r\geqslant 3$, then the differential $\tilde{d}_2^{\,0,\,n-1}$ is injective.
If the groups $\widetilde{E}^{r+2,\,n-r-1}_{r}(\mathcal{E})$ are trivial for $r\geqslant 2$, the differential $\tilde{d}_2^{\,0,\,n-1}$ is surjective.
Hence, if the groups $\widetilde{E}^{r,\,n-r}_r(\mathcal{E})$, $\widetilde{E}^{\,4,\,n-3}_{2}(\mathcal{E})$, $\widetilde{E}^{r+2,\,n-r-1}_{r}(\mathcal{E})$ are trivial for $r\geqslant 3$, we have the isomorphism
$$
\ker \tilde{d}^{\,0,\,n-1}_1 \cong \widetilde{E}_2^{2,\,n-2}(\mathcal{E}) = E^{3,\,n-2}_2(\mathcal{E})
$$
and hidden Lagrangians of $\mathcal{E}$ can be identified with elements of the group $E^{3,\,n-2}_2(\mathcal{E})$.\\[1 ex]
\definitiona{A presymplectic structure of a system $\mathcal{E}$ is \textit{extendable} if it is generated by some internal Lagrangian.}\\[-1.5ex]

If the groups
$\widetilde{E}^{r+1,\,n-r}_r(\mathcal{E})$ are trivial for $r\geqslant 3$, then the group $\ker \tilde{d}^{\,1,\,n-1}_2$ is also trivial. In this case 
the mapping 
$$
\tilde{d}^{\,1,\,n-1}_2\colon\, \widetilde{E}^{1,\,n-1}_2(\mathcal{E})\to E^{4,\,n-2}_2(\mathcal{E})
$$ 
is injective. Here the group $\widetilde{E}^{1,\,n-1}_2(\mathcal{E})$ consists of non-extendable presymplectic structures modulo extendable ones.
Combining this result with the previous one, we get the following\\[1 ex]
\theorema{\label{theor3} 
Let $a_r$ be a sequence such that $a_1 = 2$, $a_{r+1} = a_r + r$.
Suppose $\mathcal{E}$ is an infinitely prolonged system of equations such that the groups $E^{4,\, n-3}_2(\mathcal{E})$, $E^{5,\, n-3}_2(\mathcal{E})$,
$$
E^{p,\,n+1-p}_r(\mathcal{E}),\quad E^{p,\,n+2-p}_r(\mathcal{E})\qquad \text{for}\quad r\geqslant 3,\quad a_r \leqslant p < a_{r+1}
$$
of the Vinogradov $\mathcal{C}$-spectral sequence are trivial.
Then\\
1) $\ker \tilde{d}^{\,0,\,n-1}_1 \cong E^{3,\,n-2}_2(\mathcal{E})$ and hidden Lagrangians of $\mathcal{E}$ can be identified with elements of the group $E^{3,\,n-2}_2(\mathcal{E})$;\\
2) the mapping from the group of classes of non-extendable presymplectic structures
$$
\tilde{d}^{\,1,\,n-1}_2\colon\, \widetilde{E}^{1,\,n-1}_2(\mathcal{E})\to E^{4,\,n-2}_2(\mathcal{E})
$$
is injective.}\\[1 ex]
Therefore, the description of internal Lagrangians of $l$-normal differential equations~\cite{VinKr} completely amounts to presymplectic structures. The same situation occurs for all ordinary differential equations.

Let us note that, according to the $k$-line theorem~\cite{Verb}, if a compatibility complex of a system of differential equations is of length three, the assumptions of Theorem~\ref{theor3} hold. Thus, we obtain\\[1.5 ex]
\corollarya{\label{corol} Assume that an infinitely prolonged system of equations $\mathcal{E}$ admits a compatibility complex of length three. Then the group of hidden Lagrangians of $\mathcal{E}$ is isomorphic to the group $E^{3,\,n-2}_2(\mathcal{E})$ of the Vinogradov $\mathcal{C}$-spectral sequence.}\\[1 ex]
In particular, \textit{irreducible gauge theories are of this type.} The assumptions of Theorem~\ref{theor3} also hold for all systems of equations with two independent variables.\\[1.5ex]
\remarka{If the assumptions of Corollary~\ref{corol} hold, then the group $E_2^{2,\,n-2}(\mathcal{E})$ of Vinogradov's $\mathcal{C}$-spectral sequence is isomorphic to $E^{\ \!\!0,\,n-1}_2(\mathcal{E}) = \ker d_1^{\,0,\,n-1}$ (this is the group of conservation laws with trivial cosymmetries).}

\section{\label{Dis} Discussion\\[-5ex]}

\

Let $\mathcal{E}$ be an infinitely prolonged system of Euler-Lagrange equations of a certain Lagrangian $[L]_h$. Assume that $\mathcal{E}$ also possesses a hidden Lagrangian. Then we get several internal Lagrangians determining the same symplectic structure. Thus the symplectic structure does not allow one to distinguish the internal Lagrangian generated by the horizontal form $[L]_h$ from some other internal Lagrangian. The situation is additionally complicated by the fact that hidden Lagrangians are \textit{invariant under the action of any symmetry} (by means of the Lie derivative). This applies to gauge symmetries as well. It would be interesting to find an example of such a system $\mathcal{E}$ among physically significant equations. However, even for conservation laws, a similar problem is non-trivial (see~\cite{Olver}) and apparently has not yet been solved. Nevertheless, for the intrinsic description of variational principles, it seems more preferable to use the concept of an internal Lagrangian rather than the concept of a presymplectic structure. The question of whether there are hidden Lagrangians or non-extendable presymplectic structures deserves special attention.

Also note that, using internal Lagrangians of a system of differential equations, one can construct conservation laws of the corresponding prime system~\cite{Druzhkov}. Let us recall that conservation laws allow one to construct differential coverings~\cite{VinKr}. It may be that different internal Lagrangians of a system of equations define non-equivalent differential coverings over the corresponding prime system. This will be considered elsewhere.

\subsection{Abelian $p$-form theories}
Finally, let us consider the Abelian $p$-form theories.\\

Let $M$ be a (pseudo-)Riemannian manifold, $\pi\colon E \to M$ be the $p$-th exterior power of the cotangent bundle over $M$.
Assume that $\dim M = n > 2$, \ $1 \leqslant p\leqslant n-1$ and that $M$ is topologically trivial. Consider the equation
\begin{equation}
d\ast dA = 0
\label{exa}
\end{equation}
for an unknown differential $p$-form $A\in \Lambda^p(M)$.

The terms $E_1^{i,\, q}$ of the infinite prolongation of the equation~\eqref{exa} were calculated in~\cite{Verb} (for $i > 0$, $q\leqslant n-2$). Based on the arguments given there, we can conclude that the corresponding terms $E^{*,\,q}_2$ are trivial for $q \leqslant n-2$, except for $E^{0,\, 0} _2$.
Hence, the equation under consideration does not admit hidden Lagrangians. It also does not admit non-extendable presymplectic structures. Both of these conclusions immediately follow from Theorem~\ref{theor3}.\\

\centerline{\bf Acknowledgments}

\
\\[-1ex]
The author is grateful to I.S.~Krasil'shchik, A.M.~Verbovetsky, M.~Grigoriev and V.~Gritzaenko for constructive discussions.

\end{document}